\documentclass[aps,prl,showpacs,twocolumn,floatfix,amssymb,amsmath,secnumarabic,nofootinbib,superscriptaddress]{revtex4-1}

\usepackage{ctable}
\usepackage{multirow}
\usepackage{isomath}

\def\ben{\begin{equation}}
\def\een{\end{equation}}
\def\sss{\scriptscriptstyle\rm}
\def\xc{_{\sss XC}}

\def\loc{^{\rm loc}}

\def\bn{\vectorsym{n}}
\def\bk{\vectorsym{k}}
\def\bv{\vectorsym{v}}
\def\bx{\vectorsym{x}}
\def\T{\vectorsym{T}}
\def\ba{\vectorsym{\alpha}}
\def\TML{T^{\rm ML}}
\def\K{\matrixsym{K}}
\def\X{\matrixsym{X}}
\def\Vm{\matrixsym{V}}
\def\C{\matrixsym{C}}
\def\Pm{\matrixsym{P}}

\def\W{{\rm ^W}}

\begin{document}

\title{Finding Density Functionals with Machine Learning}

\author{John C. Snyder}
\affiliation{Departments of Chemistry and of Physics,
University of California, Irvine, CA 92697, USA}

\author{Matthias Rupp}
\affiliation{Machine Learning Group, Technical University of Berlin, 10587 Berlin, Germany}
\affiliation{Institute of Pharmaceutical Sciences, ETH Zurich, 8093 Z{\"u}rich, Switzerland}

\author{Katja Hansen}
\affiliation{Machine Learning Group, Technical University of Berlin, 10587 Berlin, Germany}

\author{Klaus-Robert M{\"u}ller}
\affiliation{Machine Learning Group, Technical University of Berlin, 10587 Berlin, Germany}

\author{Kieron Burke}
\affiliation{Departments of Chemistry and of Physics,
University of California, Irvine, CA 92697, USA}

\date{\today}

%%%%%%%%%
\begin{abstract}
Machine learning is used to approximate density functionals.
For the model problem of the kinetic energy of non-interacting fermions in 1d, mean absolute errors below 1 kcal/mol
on test densities similar to the training set are reached with fewer than 100 training densities.
A predictor identifies if a test density is within the interpolation region.
Via principal component analysis, a projected functional derivative finds highly accurate self-consistent densities.
Challenges for application of our method to real electronic structure problems are discussed.
\end{abstract}


%%%%%%%%%
\maketitle

%%%%%%%%%
Each year, more than 10,000 papers report solutions to electronic structure
problems using Kohn-Sham (KS) density functional theory (DFT) \cite{HK64, KS65}.
All approximate
the exchange-correlation (XC) energy as a functional of the electronic
spin densities. The quality of the results depends crucially on these density
functional approximations. For example,
present approximations often fail for strongly correlated systems, rendering
the methodology useless for some of the most interesting problems.

Thus, there is a never-ending search for improved XC approximations.  The original
local density approximation (LDA) of Kohn and Sham \cite{KS65} is uniquely defined by the
properties of the uniform gas, and has been argued to be a  universal limit of
{\em all} systems \cite{ELCB08}. But the refinements that have proven useful in 
chemistry \cite{SDCF94} and materials \cite{PBE96} are not, and differ both in their derivations and
details.   Traditionally, physicists favor a non-empirical approach,
deriving approximations from quantum mechanics and avoiding fitting
to specific finite systems \cite{PA10}.
Such non-empirical functionals can be considered controlled extrapolations that 
work well across a broad range of systems and properties, bridging
the divide between molecules and solids.
Chemists typically use a few \cite{B88, LYP88} or several
dozen \cite{ZT08} parameters to improve accuracy on a limited class of molecules.  
Empirical functionals are
limited interpolations that are more accurate for the molecular systems they 
are fitted to, but often fail for solids.
Passionate debates are fueled by this cultural divide.

{\em Machine learning} (ML) is a powerful tool for finding patterns in high-dimensional data.
ML employs algorithms by which the computer learns from empirical data via induction, and
has been very successful in many applications \cite{mmrts2001, i2007, RTML12}.
In ML,  intuition is used to choose the basic mechanism and representation of the data,
but not directly applied to the details of the model.
Mean errors can be systematically decreased with increasing number of inputs.
In contrast, human-designed empirical approximations employ standard forms 
derived from general principles, fitting the parameters to training sets.
These explore only an infinitesimal fraction of all possible
functionals and use relatively few data points.

DFT works for electronic structure because the underlying
many-body Hamiltonian is simple, while accurate solution of the
Schr\"odinger equation is very demanding.  All electrons Coulomb repel
one-another, have spin $1/2$, and are Coulomb attracted to the nuclei.  ML is a natural tool for taking maximum
advantage of this simplicity.

Here, we adapt ML to a prototype density functional
problem: non-interacting spinless fermions confined to a 1d box, subject to a smooth potential.
We define key technical concepts needed to apply ML to DFT problems.  
The accuracy we achieve in approximating the kinetic energy (KE) of this system is
far beyond the capabilities of 
any present approximations and is even sufficient
to produce highly accurate self-consistent densities.
Our ML approximation (MLA) achieves
chemical accuracy using many more inputs, but requires far less insight into the underlying physics.

\begin{figure}[htbp]
\unitlength1cm
\begin{picture}(7.6,5)
\put(0.15,0.2){\includegraphics[width=7.6cm]{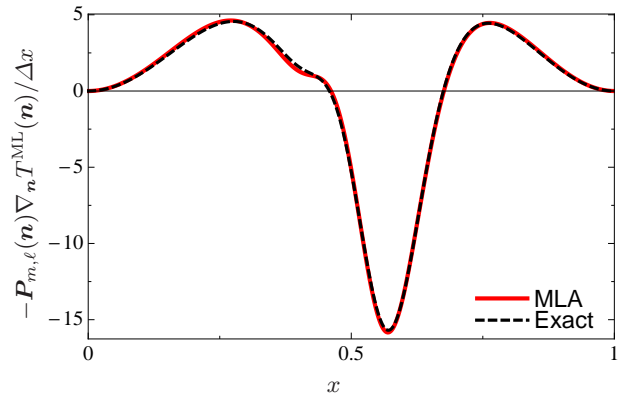}}
\put(0.15,-0.1){\makebox(7.6,0){$x$}}
\put(-0.15,0.2){\rotatebox{90}{\makebox(4.8,0){$-\Pm_{m, \ell}(\bn)\nabla_\bn \TML(\bn)/\Delta x$}}}
\end{picture}
\caption{(color online). Comparison of a projected (see within) functional derivative of our MLA with the exact curve.}
\label{fig:MLprojfuncderiv}
\end{figure}

We illustrate the accuracy of our MLA with Fig.~\ref{fig:MLprojfuncderiv},
in which the functional was constructed from 100 densities on a dense grid.
This success opens up a new approach to functional
approximation, entirely distinct from previous approaches: Our MLA contains 
$\sim10^5$ empirical numbers and satisfies none of the standard exact conditions.

The prototype DFT problem we consider is $N$ non-interacting spinless fermions confined to
a 1d box, $0 \leq x \leq 1$, with hard walls.  For continuous potentials $v(x)$, we solve the 
Schr\"odinger equation numerically with the lowest $N$ orbitals occupied, finding
the KE and the electronic density $n(x)$, the sum of the squares of the occupied orbitals.
Our aim is to construct an MLA for the KE $T[n]$ that bypasses the need to solve 
the Schr\"odinger equation---a 1d analog of orbital-free DFT \cite{KJTH09}.
(In (3d) orbital-free DFT, the local approximation, as used in Thomas-Fermi theory, is typically
accurate to within 10\%, and the addition of the leading gradient correction reduces
the error to about 1\% \cite{DG90}.  Even this small an error in the total KE
is too large to give accurate chemical properties.)

\begin{figure}[t]
\unitlength1cm
\begin{picture}(7.6,5)
\put(0.15,0.2){\includegraphics[width=7.6cm]{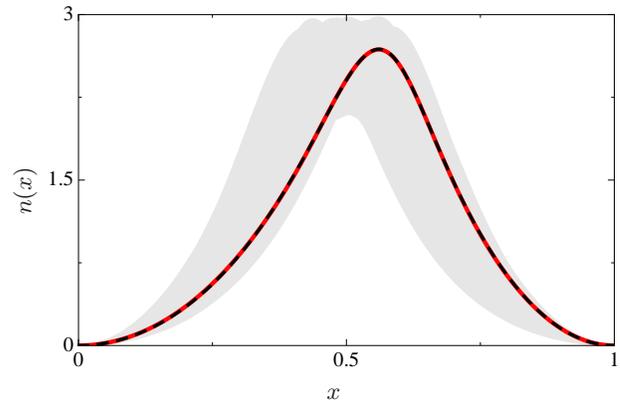}}
\put(0.15,-0.1){\makebox(7.6,0){$x$}}
\put(-0.15,0.2){\rotatebox{90}{\makebox(4.8,0){$n(x)$}}}
\end{picture}
\caption{(color online). The shaded region shows the extent of variation of $n(x)$ within our dataset for $N=1$.
Exact (red) and a self-consistent (black, dashed) density for potential of Fig.~\ref{fig:MLwigglyfuncderiv}.}
\label{fig:MLdens}
\end{figure}

First, we specify a class of potentials from which we generate densities,
which are then discretized on a uniform grid of $G$ points.
We use a linear combination of 3 Gaussian dips with different depths, widths,
and centers:
\ben
v(x) = -\sum_{i=1}^3 a_i \exp(-(x-b_i)^2/(2c_i^2)).
\een
We generate 2000 such potentials, randomly sampling  $ 1 < a < 10$, $ 0.4 < b < 0.6$, and
$ 0.03 < c < 0.1$.
For each $v_j(x)$, we find, for $N$ up to 4 electrons,
the KE $T_{j,N}$ and density $\bn_{j,N} \in \mathbb{R}^G$
on the grid using Numerov's method \cite{N24}.
For $G=500$, the error in $T_{j,N}$ due to
discretization is less than $1.5 \times 10^{-7}$.
We take 1000 densities as a 
{\em test set}, and choose $M$ others for training. The
variation in this dataset for $N=1$ is illustrated in Fig. \ref{fig:MLdens}.

Kernel ridge regression (KRR) is a non-linear version of 
regression with regularization to prevent overfitting \cite{htf2009}.
For KRR, our MLA takes the form
\ben
\TML(\bn) = \bar T \sum_{j=1}^M \alpha_j k(\bn_j, \bn),
\label{eq:TMLdef}
\een
where $\alpha_j$ are weights to be determined, $\bn_j$ are training densities and $k$ is the kernel,
which measures similarity between densities. Here $\bar T$ is the mean KE of the training set, inserted
for convenience.
We choose a Gaussian kernel, common in ML:
\ben
k(\bn, \bn') = \exp(-\| \bn - \bn' \|^2/(2\sigma^2)),
\een
where the hyperparameter $\sigma$ is called the length scale.
The weights are found by minimizing the cost function
\ben
{\mathcal C}(\ba) =  \sum_{j=1}^M \Delta T_j ^2 +  \lambda \| {\ba} \|^2,
\een
where $\Delta T_j = \TML_j - T_j$ and $\ba = (\alpha_1, \dots, \alpha_M)$.
The second term is a regularizer that
penalizes large weights to prevent overfitting.
The hyperparameter $\lambda$ controls regularization strength. 
Minimizing ${\mathcal C}(\ba)$ gives
\ben
\ba = (\K + \lambda \matrixsym{I})^{-1} \T,
\label{eq:KRRsolution}
\een
where $\K$ is the kernel matrix, with elements
$\K_{ij} = k(\bn_i, \bn_j)$, and $\matrixsym{I}$ is the identity matrix.
Then $\sigma$ and $\lambda$ are determined through 10-fold cross-validation:
The training set is partitioned into 10 bins of equal size. For each bin, the functional is trained on the remaining samples
and $\sigma$ and $\lambda$ are optimized by minimizing the mean absolute error (MAE) on
the bin. The partitioning is repeated up to 40 times and the hyperparameters are chosen as the median over all bins.

\renewcommand*\arraystretch{1.4}
\begin{table}[htbp]
	\begin{tabular}{c@{\hspace{0.8em}}c@{\hspace{0.8em}}c@{\hspace{0.6em}}r@{\hspace{1.6em}}l@{\hspace{1em}}l@{\hspace{1em}}l}
	\toprule[0.1em]
		$N$ & $M$ & $\lambda\times 10^{14}$ & $\sigma$ & $\overline{| \Delta T |}$ & $| \Delta T |^{\text{std}}$ & $| \Delta T |^{\text{max}}$ \\
	\toprule[0.1em]
	\multirow{6}*{1}
		& 40   	& $57600$ 	& 238 	& 3.3		& 3.0		& 23  \\
		& 60		& $10000$ 	& 95 	& 1.2		& 1.2		& 10 \\
		& 80		& $4489$ 		& 48 	& 0.43		& 0.54		& 7.1 \\
		& 100   	& $12$ 	& 43		& 0.15[3.0]	& 0.24[5.3]	& 3.2[46]  \\
		& 150	& $6.3$ 		& 33 	& 0.06		& 0.10		& 1.3 \\
		& 200	& $3.2$ 		& 28 	& 0.03		& 0.05		& 0.65 \\
	\midrule[0.05em]
		2 & 100	& $1.7$ 		& 52 	& 0.13[1.4] 	& 0.20[3.0] 	& 1.8[37] \\
		3 & 100	& $4.0$ 		& 74 	& 0.12[0.9]	& 0.18[1.5]	& 1.8[14] \\
		4 & 100	& $2.0$ 		& 73 	& 0.08[0.6]	& 0.14[0.8]	& 2.3[6] \\
	\midrule[0.05em]
		1-4$^\dagger$ & 400 & $3.2$ & 47 & 0.12	& 0.20 		& 3.6 \\ 
	\bottomrule[0.1em]
	\end{tabular}
	\caption{Parameters and errors (mean absolute, std. dev., and max abs. in kcal/mol)
	as a function of electron number $N$ and number of training densities $M$. Brackets
	represent errors on self-consistent densities with $m=30$ and $\ell = 5$. The $\alpha_j$ are on the order of $10^6$ 
	and both positive and negative \cite{suppinfo}.
	$^\dagger $Training set includes $\bn_{j,N}$, for $j=1,\dots,100$, $N=1,\dots,4$.}
	\label{tbl:modelerrors}
\end{table}

Table \ref{tbl:modelerrors} gives the performance of $\TML$ (Eq. \ref{eq:TMLdef})
trained on $M$ $N$-electron densities and evaluated on the corresponding test set.
The mean KE of the test set for $N=1$ is 5.40 Hartree (3390 kcal/mol).  
To contrast, the LDA in 1d is
$T\loc[n] = \pi^2 \int dx\, n^3(x)/6$
and the von Weizs\"acker functional is
$T\W[n]=\int dx\, n'(x)^2/(8n(x))$.
For $N=1$, the MAE of $T\loc$ on the test set is 217 kcal/mol and the modified gradient
expansion approximation \cite{LCPB09},
$T^{\rm MGEA}[n] = T\loc[n] - c\, T\W[n]$,
has a MAE of 160 kcal/mol, where $c=0.0543$ has been chosen to minimize the error (the gradient correction
is not as beneficial in 1d as in 3d).
For $\TML$, both the mean and maximum absolute errors improve as $N$ or $M$ increases (the system
becomes more uniform as $N\to\infty$ \cite{ELCB08}).
At $M=80$, we have already achieved ``chemical accuracy,'' i.e., a MAE below 1 kcal/mol.
At $M=200$, {\em no} error is above 1 kcal/mol.
Simultaneously incorporating different $N$ into the training set has little effect on the overall performance.

\begin{figure}[t]
\unitlength1cm
\begin{picture}(7.6,5)
\put(0.15,0.2){\includegraphics[width=7.6cm]{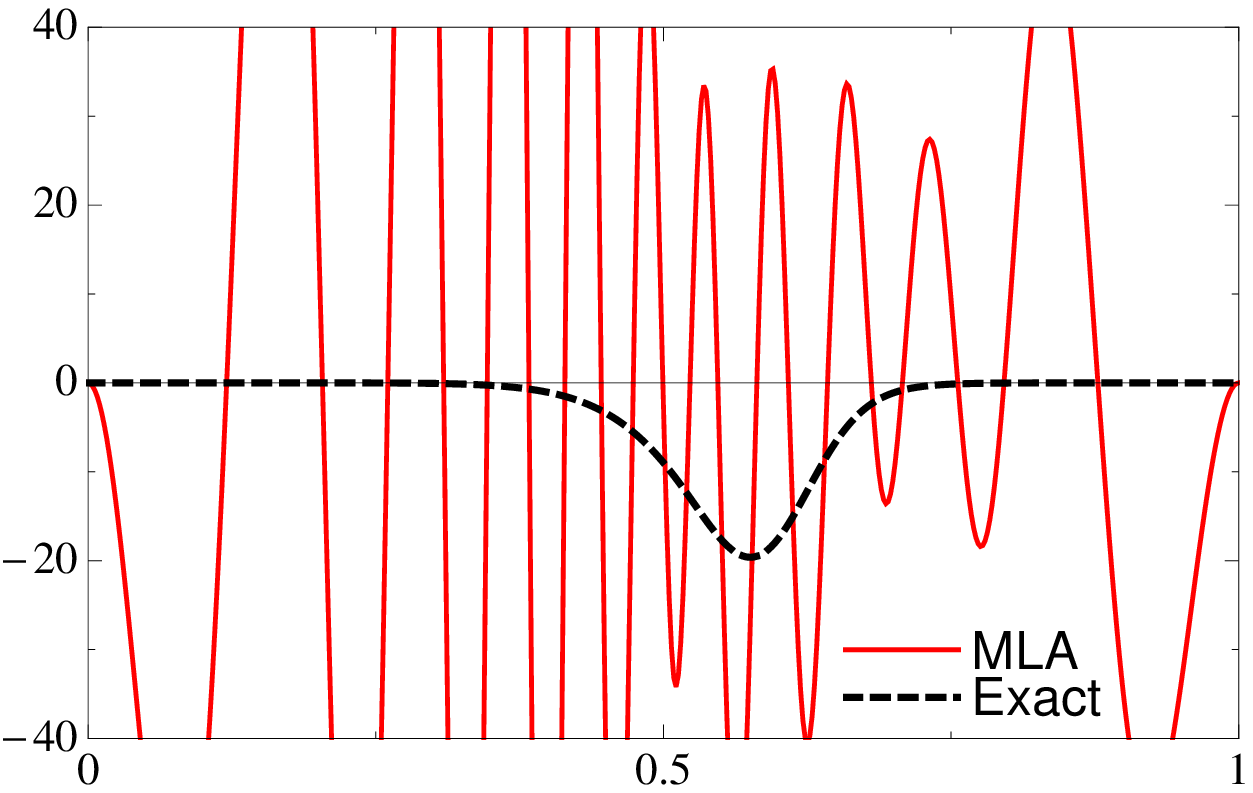}}
\put(0.15,-0.1){\makebox(7.6,0){$x$}}
\put(-0.15,0.2){\rotatebox{90}{\makebox(4.8,0){$-\nabla_\bn \TML(\bn)/\Delta x$}}}
\end{picture}
\caption{(color online). Functional derivative of $\TML$, evaluated on the density of Fig.~\ref{fig:MLdens}.}
\label{fig:MLwigglyfuncderiv}
\end{figure}

With such unheard of accuracy, it is tempting to declare ``mission accomplished,'' but this would
be premature.  A KE functional that predicts only the energy is useless in practice, since orbital-free
DFT uses
functional derivatives in self-consistent procedures to {\em find} the density within a given
approximation, via
\ben
\frac{\delta T[n]}{\delta n(x)} = \mu- v(x),
\een
where $\mu$ is adjusted to produce the required particle number.
The (discretized) functional derivative of $\TML$ is
\ben
\frac{1}{\Delta x} \nabla_\bn \TML(\bn) = \sum_{j=1}^M \alpha_j' (\bn_j - \bn) k(\bn_j, \bn),
\een
where $\alpha_j' = \alpha_j/(\sigma^2 \Delta x)$.
This oscillates wildly relative to the exact curve (Fig.~\ref{fig:MLwigglyfuncderiv}),
typical behavior that does {\em not} improve with increasing $M$.
No finite interpolation can accurately reproduce all details of a functional derivative.

We overcome this problem using principal component analysis (PCA).
The space of all densities is contained in $\mathbb{R}^G$, but only a few directions in this
space are relevant.  For a given density $\bn$, find the $m$
training densities $(\bn_{j_1}, \dots, \bn_{j_m})$
closest to $\bn$. Construct the covariance matrix of directions from $\bn$ to each training density
$\C = \X^\top \X/m,$
where $\X = (\bn_{j_1} - \bn, \dots, \bn_{j_m} - \bn)^\top$. 
Diagonalizing $\C \in \mathbb{R}^{G \times G}$ gives eigenvalues $\lambda_j$ and
eigenvectors $\bx_j$ which we list in decreasing order.
The $\bx_j$ with larger $\lambda_j$ are directions with substantial variation in the dataset.
Those with $\lambda_j$ below a cutoff are irrelevant \cite{suppinfo}. 
In these extraneous dimensions, there is too little variation within the dataset, producing 
noise in the model functional derivative. By projecting onto the subspace spanned by the
relevant dimensions, we eliminate this noise. This projection is given by
$\Pm_{m, \ell}(\bn) = \Vm^\top \Vm,$
where $\Vm = (\bx_1, \dots, \bx_\ell)^\top$ and $\ell$ is the number of relevant eigenvectors.
In Fig \ref{fig:MLprojfuncderiv}, with $m=30$ and $\ell=5$, the projected functional
derivatives are in excellent agreement.

The ultimate test for a density functional is to produce a self-con\-sis\-tent density that minimizes the total energy and check its error.
This error will be several times larger than that of the functional evaluated on the exact
density.  For example, $T\loc$ on particles in 1d flat boxes always gives 4 times larger error.
To find a minimizing density, perform a gradient descent search restricted to the 
local PCA subspace:
Starting from a guess $\bn^{(0)}$, take a small step in the
opposite direction of the projected functional derivative of the total energy in each iteration $j$:
\ben
\bn^{(j+1)} = \bn^{(j)} - \epsilon \Pm_{m, \ell}(\bn^{(j)})(\bv + \nabla_\bn \TML(\bn^{(j)})/\Delta x),
\label{eq:gradsearch}
\een
where $\epsilon$ is a small number and $\bv$ is the discretized potential.
The search is unstable if $\ell$ is too large, inaccurate if $\ell$ is too small, and relatively insensitive to $m$ \cite{suppinfo}.

The performance of $\TML$ in finding self-consistent densities is given in Table \ref{tbl:modelerrors}.
Errors are an order of magnitude larger than
that of $\TML$ on the exact densities.  We do not find a unique density,
but instead a set of similar densities depending on the initial guess (e.g. Fig. \ref{fig:MLdens}). 
The density with lowest total energy
 does {\em not} have the smallest error.  Although the search does not
produce a unique minimum, it produces a range of similar but valid approximate
densities, each with a small error. 
Even with an order of magnitude larger error, we still reach chemical accuracy, now on self-consistent
densities.
No existing KE approximation comes close to this performance.

\begin{figure}[htbp]
\unitlength1cm
\begin{picture}(7.6,5.1)
\put(0.15,0.3){\includegraphics[width=7.6cm]{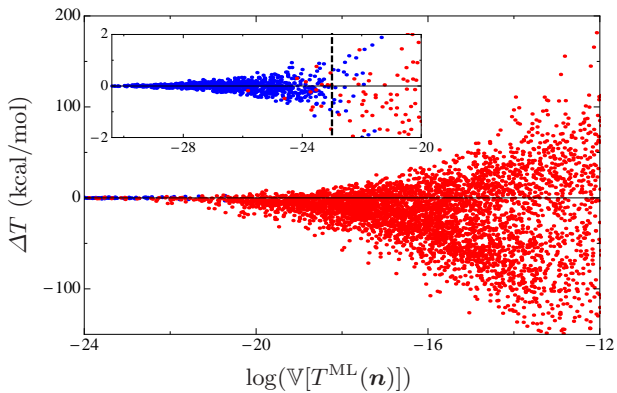}}
\put(0.15,0){\makebox(7.6,0){$\log(\mathbb{V}[\TML(\bn)])$}}
\put(-0.15,0.3){\rotatebox{90}{\makebox(4.8,0){$\Delta T$ (kcal/mol)}}}
\end{picture}
\caption{(color online). The correlation between MLA error and predictive variance for $N=1$, $M=100$. Each point represents a density 
in the test set (blue) or new dataset (red). The vertical line denotes the transition between interpolation and extrapolation.}
\label{errvspredvar}
\end{figure}

What are the limitations of this approach?  ML is a balanced interpolation on known data, and
should be unreliable for densities far from the training set.  To demonstrate this, we generate a new dataset of 5000
densities with $N=1$ for an expanded parameter range:
$0.1 < a < 20$, $0.2 < b < 0.8$ and $0.01 < c < 0.3$.
The predictive variance (borrowed from Gaussian process regression \cite{rw2006})
\ben
\mathbb{V}[\TML(\bn)] = k(\bn, \bn) - \bk(\bn)^\top (\K + \lambda \matrixsym{I})^{-1} \bk(\bn),
\een
where $\bk(\bn) = (k(\bn_1, \bn), \dots, k(\bn_M, \bn))$, is a measure of the uncertainty in the prediction $\TML(\bn)$ due to 
 sparseness of training densities around $\bn$. 
In Fig. \ref{errvspredvar}, we plot the error $\Delta T$ as a function of $\log(\mathbb{V}[\TML(\bn)] )$, for both the
test set and the new dataset, showing a clear correlation.
From the inset, we expect our MLA to deliver chemical accuracy for $\log(\mathbb{V}[\TML(\bn)]) < -24$.

Does ML allow for human intuition?  In fact, the more prior
knowledge we insert into the MLA, the higher the accuracy we can achieve.  Writing 
$T = T\W + T_\theta $,
where $T_\theta \geq 0$ \cite{KJTH09}, we 
repeat our calculations to find an MLA for $T_\theta$.  For $N=1$ we get almost zero error,
and a factor of 2-4 reduction of error otherwise.  Thus, intuition about the functional
can be built in to improve results.

The primary interest in KS DFT is XC for molecules and solids.  We have
far less information about this than in  the
prototype studied here.  For small molecules and simple solids,
direct solutions of the Schr\"odinger equation yield highly accurate
values of $E\xc$.  Imagine a sequence of models, beginning with atoms,
diatomics, etc., in which such accurate results are used as training data for an MLA.
Key issues are how accurate a functional can be attained with a finite number of data,
and what fraction of the density space it is accurate for.

A more immediate target is the non-interacting KE in 
KS DFT calculations.  An accurate approximation would allow finding
densities and energies {\em without} solving the KS equations, greatly increasing the
speed of large calculations \cite{KJTH09}.
The key differences
with our prototype is the three-dimensional nature, the Coulomb singularities, and
the variation with nuclear positions.  For this problem,
finding self-consistent densities is crucial, and hence our focus here.
But in the 3d case, {\em every} KS calculation ever run, including every iteration
in a self-consistent loop, generates training data---a density, KE,
KS potential and functional derivative.  
The space of systems, including both solids and molecules, is vast, but could be approached in small steps.

Two last points:
The first is that this type of empiricism is qualitatively distinct from that
present in the literature.  The choices made are those customary in ML, and
require no intuition about the physical nature of the problem.
Second, the approximation is expressed in terms of about $10^5$ numbers, and only the projected functional
derivative is accurate.  We have no simple way of comparing such approximations to those presently
popular.  
For example, for $N=1$ in the prototype, the exact functional is $T\W$. How is this related to our MLA, and how does our
MLA account for this exact limit?

The authors thank the Institute for Pure and Applied Mathematics
at UCLA for hospitality and acknowledge NSF CHE-1112442 (JS, KB), 
EU PASCAL2 and DFG MU 987/4-2 (MR, KH, KRM) and EU Marie Curie
IEF 273039 (MR).

%%%%%%%%%

%%%%%%%%%
\label{page:end}

\begin{thebibliography}{99}

\bibitem{HK64}
P. Hohenberg and W. Kohn, Phys. Rev. B {\bf 136}, 864 (1964).

\bibitem{KS65}
W. Kohn and L. J. Sham, Phys. Rev. A {\bf 140}, 1133 (1965).

\bibitem{ELCB08}
P. Elliott, D. Lee, A. Cangi, and K. Burke, Phys. Rev. Lett. {\bf 100}, 256406 (2008).

\bibitem{SDCF94}
P. J. Stephens, F. J. Devlin, C. F. Chabalowski, and M. J. Frisch, J. Phys. Chem. {\bf 98}, 11623 (1994).

\bibitem{PBE96}
J. P. Perdew, K. Burke, and M. Ernzerhof, Phys. Rev. Lett. {\bf 77}, 3865 (1996).

\bibitem{PA10}
J. P. Perdew and A. Ruzsinszky, Int. J. Quant. Chem. {\bf 110}, 2801 (2010).

\bibitem{B88}
A. D. Becke, Phys. Rev. A {\bf 38}, 3098 (1988).

\bibitem{LYP88}
C. Lee, W. Yang, and R. G. Parr, Phys. Rev. B {\bf 37}, 785 (1988).

\bibitem{ZT08}
Y. Zhao and D. Truhlar, Theor. Chem. Accounts {\bf 120}, 215 (2008).

\bibitem{mmrts2001}
K.-R. M\"uller, S. Mika, G. R\"atsch, K. Tsuda, and B. Sch\"olkopf, IEEE Trans. Neural Network {\bf 12}, 181 (2001).

\bibitem{i2007}
O. Ivanciuc, in {\em Reviews in Computational Chemistry}, edited by K. Lipkowitz and T. Cundari (Wiley, Hoboken, 2007), Vol. 23, p. 291.

\bibitem{RTML12}
M. Rupp, A. Tkatchenko, K.-R. M{\"u}ller, and O. A. von Lilienfeld, Phys. Rev. Lett. (to be published).

\bibitem{KJTH09}
V. Karasiev, R. Jones, S. Trickey, and F. Harris, in {\em New Developments in Quantum Chemistry}, edited by J. Paz
and A. Hern\'andez (Research Signpost, Kerala, in press).

\bibitem{DG90}
R. M. Dreizler and E. K. U. Gross, {\em Density Functional Theory: An Approach to the Quantum Many-Body Problem} (Springer, Berlin, 1990).

\bibitem{N24}
See e.g. E. Hairer, P. N{\o}rsett, P. Syvert Paul and G. Wanner, {\em Solving ordinary differential equations I: Nonstiff problems} (Springer, New York, 1993).
%B. V. Numerov, Roy. Astro. Soc. Monthly Notices {\bf 84}, 592 (1924).

\bibitem{htf2009}
T. Hastie, R. Tibshirani, and J. Friedman, {\em The Elements of Statistical Learning. Data Mining, Inference, and Prediction}, 2nd ed. (Springer, New York, 2009).

\bibitem{suppinfo}
See Supplemental Material (appended at the end of this manuscript) for information necessary to construct the MLA functional and more detail on the PCA projections and self-consistent densities.

\bibitem{LCPB09}
D. Lee, L. A. Constantin, J. P. Perdew, and K. Burke, J. Chem. Phys. {\bf 130}, 034107 (2009).

\bibitem{rw2006}
C. Rasmussen and C. Williams, {\em Gaussian Processes for Machine Learning} (MIT Press, Cambridge, 2006).

\end{thebibliography}
\end{document}

\title{Supplementary Information for ``Finding Density Functionals with Machine Learning''}

\author{John C. Snyder}
\affiliation{Departments of Chemistry and of Physics,
University of California, Irvine, CA 92697, USA}

\author{Matthias Rupp}
\affiliation{Machine Learning Group, Technical University of Berlin, 10587 Berlin, Germany}
\affiliation{Institute of Pharmaceutical Sciences, ETH Zurich, 8093 Z{\"u}rich, Switzerland}

\author{Katja Hansen}
\affiliation{Machine Learning Group, Technical University of Berlin, 10587 Berlin, Germany}

\author{Klaus-Robert M{\"u}ller}
\affiliation{Machine Learning Group, Technical University of Berlin, 10587 Berlin, Germany}

\author{Kieron Burke}
\affiliation{Departments of Chemistry and of Physics,
University of California, Irvine, CA 92697, USA}

\date{\today}

%%%%%%%%%
\maketitle

%%%%%%%%%

The purpose of this supplementary material is to provide the information necessary to
reproduce our machine learning approximation (MLA) to the kinetic energy functional exactly,
as well as extra details about the nature of the MLA. Table \ref{tbl:MLAfunc} gives
the first 100 potential parameters $a_i$, $b_i$, and $c_i$ to double precision. Along with the
equations in the main paper and details of Numerov's method, one should be able to reproduce 
the corresponding weights $a_j$ of the functional on line 4 of Table \uppercase\expandafter{\romannumeral 1\relax} in the main text.

%%%%%%%%%
\section*{Self-consistent densities}

Table \ref{tbl:PCAevs} gives the percentage of variance lost, $(1-\sum_{j=1}^\ell \lambda_j / \sum_{j=1}^G \lambda_j) \times 100\%$, in taking the first $\ell$ eigenvalues in the principle component analysis (PCA) projection.  For each projection, there are $G=500$ eigenvalues, but only a few are necessary
to represent the local variation in the densities. For example, with $N=1$, $m=30$ and $\ell=5$, we retain 99.98\% of the variance when projecting onto the PCA subspace.
In Table \ref{tbl:scdparamopt}, we coarsely optimize the parameters $m$ and $\ell$ in the PCA projection.

%%%%%%%%%
\renewcommand*\arraystretch{1.5}
\begin{table}[htbp]
	\begin{tabular}{c@{\hspace{0.8em}}D{,}{}{2.2}D{,}{}{1.1}D{,}{.}{2.2}D{,}{.}{1.2}D{,}{.}{2.3}D{,}{.}{1.4}D{,}{.}{1.4}D{,}{.}{1.4}}
	\toprule[0.1em]
		\backslashbox{$N$}{$\ell$} & \multicolumn{1}{c}{1} & \multicolumn{1}{c}{2} & \multicolumn{1}{c}{3} & \multicolumn{1}{c}{4} & \multicolumn{1}{c}{5} & \multicolumn{1}{c}{6} & \multicolumn{1}{c}{7} & \multicolumn{1}{c}{8} \\
	\toprule[0.1em]
1 & 3,5 & ,3 & 0,8 & 0,07 & 0,02 & 0,004 & 0,002 & 0,0003 \\
2 & 4,5 & 1,5 & 3,7 & 0,36 & 0,10 & 0,019 & 0,006 & 0,0015 \\
3 & 4,4 & 1,8 & 4,9 & 0,78 & 0,16 & 0,043 & 0,012 & 0,0037 \\
4 & 4,8 & 2,4 & 9,5 & 1,9 & 0,40 & 0,087 & 0,023 & 0,0063 \\
	\bottomrule[0.1em]
	\end{tabular}
	\caption{Percent of variance lost in taking the first $\ell$ PCA eigenvalues with $m=30$. Averaged over 100 PCA
	projections $\Pm_{m, \ell}(\bn)$ with randomly chosen centers $\bn$ in the test set.}
	\label{tbl:PCAevs}
\end{table}

%%%%%%%%%
\renewcommand*\arraystretch{1.5}
\begin{table}[t]
	\begin{tabular}{c@{\hspace{0.6em}}D{,}{.}{2.2}D{,}{.}{2.2}D{,}{.}{2.3}D{,}{.}{2.2}}
	\toprule[0.1em]
		\backslashbox{$m$}{$\ell$} & \multicolumn{1}{c}{3} & \multicolumn{1}{c}{4} & \multicolumn{1}{c}{5} & \multicolumn{1}{c}{6} \\
	\toprule[0.1em]
 15 & 6,5 & 1,9 & 0,92 & 1,2 \\
 20 & 6,3 & 1,9 & 0,87 & 0,89 \\
 25 & 5,2 & 1,5 & 0,87 & 0,92 \\
 30 & 4,2 & 1,3 & 0,86 & 0,97 \\
 35 & 4,6 & 1,2 & 0,90 & 1,0 \\
 40 & 3,9 & 1,2 & 0,95 & 1,0 \\
 45 & 3,9 & 1,3 & 0,98 & 0,98 \\
	\bottomrule[0.1em]
	\end{tabular}
	\caption{Mean absolute errors $\overline{| \Delta T |}$, in kcal/mol, on 100 randomly chosen self-consistent densities in the test set, averaged over $N=1,2,3$ and 4.
	This coarse optimization gives $m=30$, $\ell = 5$. Errors are less sensitive to $m$ than $\ell$. For $\ell \geq 7$, the gradient descent search fails to converge in some cases.}
	\label{tbl:scdparamopt}
\end{table}

%%%%%%%%%
\section*{Numerov's method}

\newcommand{\sgn}{\operatorname{sgn}}

We use Numerov's method \cite{N24} to solve Schr\"odinger's equation for $N$ non-interacting spinless fermions
confined to 1d box:
\ben
\left(-\frac{1}{2}\frac{\partial^2}{\partial x^2} + v(x) \right) \psi(x) = \epsilon \psi(x),
\een
with boundary conditions $\psi(0) = \psi(1) = 0$. We discretize $\psi(x)$ and $v(x)$ on a uniform grid with spacing $\Delta x = 1/(G-1)$: $\psi_j$ = $\psi(j / (G-1))$, $v_j$ = $v(j / (G-1)))$ for $j=0,\dots,G-1$.
Starting from $\psi_0 = 0$ and $\psi_1 = 1$, we calculate the remaining $\psi_j$ iteratively:
\ben
\psi_{j+1} = { (2-5\Delta x^2 f_j/6) \psi_j - (1 + \Delta x^2 f_{j-1}) /12 ) \psi_{j-1} \over  1+\Delta x^2 f_{j+1}/12 },
\een
where $f_j$ = $2(\epsilon - v_j))$. To determine the eigenvalues $\epsilon^{(k)}$ and eigenfunctions $\psi^{(k)}$,
we find the first $N$ intervals that contain a root of $\psi_{G-1}(\epsilon)$, scanning from
$\epsilon = -3\sum_{i=1}^3 a_i$ in steps of $\Delta \epsilon = 1$.
For each interval $k$, we perform a binary search for the root, reducing the length of the interval to less than $10^{-14}$.
The eigenvalue is taken as the midpoint of the interval (the error in $\epsilon_k$ is less than $5 \times 10^{-15}$).
After normalizing the eigenfunctions $\psi^{(k)}$, the density and kinetic energy are given by:
\ben
n(x_j) = \sum_{k=1}^N \psi^{(k)}(x_j)^2, \quad T = \sum_{k=1}^N \epsilon_k - \sum_{j=1}^{G-1} n(x_j) v_j \Delta x.
\een

%%%%%%%%%
\renewcommand*\arraystretch{1.1}
\begin{table}[p]
\resizebox{17.8cm}{!} {
	\begin{tabular}{lD{,}{.}{4.16}D{,}{.}{2.16}D{,}{.}{2.16}D{,}{.}{3.16}D{,}{.}{3.16}D{,}{.}{2.16}D{,}{.}{3.16}D{,}{.}{3.16}D{,}{.}{2.16}D{,}{.}{3.16}}
	\toprule[0.1em]
		$j$ & \multicolumn{1}{c}{$\alpha_j$} & \multicolumn{1}{c}{$a_1$} & \multicolumn{1}{c}{$b_1$} & \multicolumn{1}{c}{$c_1$} & \multicolumn{1}{c}{$a_2$} & \multicolumn{1}{c}{$b_2$} & \multicolumn{1}{c}{$c_2$} & \multicolumn{1}{c}{$a_3$} & \multicolumn{1}{c}{$b_3$} & \multicolumn{1}{c}{$c_3$} \\
	\toprule[0.1em]
 1 & 14,34267159730358 & 1,180270056381577 & 0,07195101071267996 & 0,5299345325943254 & 9,01320520984288 & 0,092675711955144 & 0,5396431787675333 & 7,21030205171728 & 0,0740833092851918 & 0,4120341150720884 \\
 2 & 5,175360849250056 & 6,168993733519782 & 0,082993252991643 & 0,5167227315230085 & 8,30495143619142 & 0,0890145060213322 & 0,4169122081529295 & 8,80487613349391 & 0,0965994509648079 & 0,5711057094413816 \\
 3 & 0,835534193378979 & 9,08090433047071 & 0,07685302774688097 & 0,5622686436063149 & 3,911674577648888 & 0,07551832009012722 & 0,5086649497775319 & 1,809709070880952 & 0,095976684502697 & 0,4024220371493795 \\
 4 & -5,373806223157635 & 8,69723148795899 & 0,06384871663070496 & 0,5859464293596277 & 8,67330877015531 & 0,0876928787020957 & 0,4850650257010839 & 4,189229162595343 & 0,04518368077127609 & 0,5819244871260409 \\
 5 & -0,3190730633815225 & 1,880199805158718 & 0,03425540298495952 & 0,4874044587130534 & 3,541514636747403 & 0,0983660639956121 & 0,4515151404198366 & 9,82668062409708 & 0,03264617089733192 & 0,4352642564533179 \\
 6 & 0,4351315890942081 & 3,522977833467916 & 0,07218055206599771 & 0,4489170887611125 & 3,332712157375289 & 0,07852548530734836 & 0,5852941028010116 & 8,14461268619626 & 0,04848929881736392 & 0,5484789894348707 \\
 7 & -5,783654943336363 & 5,286896436167961 & 0,04973324501652048 & 0,4653609943491444 & 7,135152688153955 & 0,0956349249199585 & 0,4408949440211174 & 2,243825016881491 & 0,05865588304821763 & 0,5687481170088626 \\
 8 & 5,857253335721763 & 1,896112874949486 & 0,03260897886438952 & 0,4501033283280557 & 5,885541925485173 & 0,06240932269593134 & 0,5618611754744713 & 6,486450948532671 & 0,07293633775145438 & 0,5695359084287562 \\
 9 & 0,3627795973240678 & 7,516612508143043 & 0,0993184090181399 & 0,5699793290550055 & 5,262683987610352 & 0,0972169632428748 & 0,4726711603381638 & 8,89032075863951 & 0,05632091198742537 & 0,4056148051043074 \\
 10 & 29,95678925033907 & 8,49309248159469 & 0,0647507853307116 & 0,4481376527388373 & 3,62263291813646 & 0,04018659271978259 & 0,5253173968297499 & 1,155756643175955 & 0,0910615024672208 & 0,5870317055827734 \\
 11 & -8,10086522849575 & 7,105439309279282 & 0,0843184446680493 & 0,4223871323591302 & 4,606936643431428 & 0,06326693058647106 & 0,5381643193286885 & 2,832289239384128 & 0,07970384658584057 & 0,4171275695759011 \\
 12 & -1,631527652578485 & 8,53955034289036 & 0,07284904722467074 & 0,5839058233689045 & 1,340024891088614 & 0,0781592051575039 & 0,5383353695983426 & 7,396341414064901 & 0,05945110164542845 & 0,4924043842188951 \\
 13 & 11,0617514474351 & 7,853417776611835 & 0,07070077317679758 & 0,5080787363762617 & 4,408481706562826 & 0,06298528328388489 & 0,4611686076617371 & 7,90542867326401 & 0,03887933278705098 & 0,5872227397532248 \\
 14 & 4,44089329011033 & 6,39698431139937 & 0,05730660668578378 & 0,5964415398755241 & 5,472334579600375 & 0,06570186368826426 & 0,5473728709183074 & 4,715521015256764 & 0,0803629674069167 & 0,5354179520994888 \\
 15 & 2,731504789238804 & 4,228544205571447 & 0,06071899009977901 & 0,4135235776817701 & 2,550224851908171 & 0,06424771468342688 & 0,5203395167113864 & 4,842862964307187 & 0,06333995603635176 & 0,5244121975227667 \\
 16 & -12,35513145514341 & 2,579686666551495 & 0,0664318905447018 & 0,4888525450630048 & 3,764107410086705 & 0,06576720199559433 & 0,5481686020053234 & 1,399253719979948 & 0,07540490302513119 & 0,4655456988965731 \\
 17 & 11,55095248921101 & 1,107700596345898 & 0,05955381913217193 & 0,5359207521608914 & 2,706093393731662 & 0,06561841006617781 & 0,4269114966922994 & 1,501048590877154 & 0,06352034496843162 & 0,4331821043592771 \\
 18 & 0,1703484854064119 & 8,26435624701031 & 0,04624196264776803 & 0,4236237608297072 & 7,025290029097274 & 0,07980750197266955 & 0,4683927184025036 & 2,822503099972245 & 0,05222522416667061 & 0,4810998521431096 \\
 19 & -0,13017855474333 & 8,74124547878586 & 0,0628793144768381 & 0,4736196423196364 & 8,59166251408243 & 0,084738261964188 & 0,4564662508099403 & 8,22877749776242 & 0,03727774324972055 & 0,4697350798193826 \\
 20 & -1,646528531854399 & 4,545173787725101 & 0,03433189693421491 & 0,4919021595234177 & 1,726935494533148 & 0,03135796595505673 & 0,580800322296872 & 2,502191185545461 & 0,0856788809687475 & 0,5909621331611888 \\
 21 & -2,371179702223466 & 1,015154584908506 & 0,07726595708511765 & 0,4008988537732911 & 6,688534202733123 & 0,0869544063742346 & 0,403293222827825 & 6,297133313147086 & 0,03345369328197418 & 0,5433070470097552 \\
 22 & 13,69383242214402 & 3,283098134816701 & 0,06841617803957927 & 0,4304195496937475 & 5,887062834756559 & 0,05152399439496422 & 0,4321995606997443 & 3,207122477750064 & 0,07371808389404321 & 0,4967284534354185 \\
 23 & -2,216419772035322 & 2,056773486230531 & 0,0825699915945694 & 0,479224783903814 & 6,945590707033352 & 0,07408636456892297 & 0,4957750456547108 & 5,846222671659907 & 0,0836452454472081 & 0,4094720200825416 \\
 24 & -6,00076435563604 & 2,03703352205463 & 0,07605517542793995 & 0,4501291515683585 & 1,45256356478288 & 0,07433692379389613 & 0,4233969076404344 & 8,95059302019488 & 0,05703615620837187 & 0,4679023659842935 \\
 25 & 8,16409108667899 & 6,400475835681034 & 0,03425589949139343 & 0,5544095818848655 & 6,086484474478635 & 0,07282844956128038 & 0,4272596897186396 & 5,36207646433345 & 0,05780634160661002 & 0,5503173198658324 \\
 26 & -22,3897314906413 & 3,751807395744921 & 0,0852624107335768 & 0,4453450848668749 & 7,559337237177321 & 0,05840909867970421 & 0,4990223467101435 & 4,729884445483556 & 0,0929218712306724 & 0,5590503374710945 \\
 27 & 5,595799173773367 & 4,522456026868127 & 0,03620201946841309 & 0,5050210060676815 & 1,885091277540955 & 0,06521780385848891 & 0,4961601035955842 & 9,47421945496329 & 0,0556740133396159 & 0,4834045749740091 \\
 28 & 1,94645582566549 & 2,751346713386582 & 0,0948195528516815 & 0,4901204212506473 & 7,863592227266858 & 0,0951797983951825 & 0,5196620049802691 & 9,51608680035275 & 0,07785377315463609 & 0,5806015592262945 \\
 29 & -1,105559822533069 & 5,725277193011525 & 0,03097303116261406 & 0,5511177785567803 & 7,988513342309234 & 0,0928323179868588 & 0,5146415116367569 & 1,101098863639624 & 0,0882939028519791 & 0,5265108634130272 \\
 30 & -3,886381532577754 & 4,95871172966177 & 0,06823254438777603 & 0,5645951313008806 & 9,08135225317634 & 0,04472856905840844 & 0,4299117124064201 & 8,24973637849809 & 0,08854431361856 & 0,5416193648861009 \\
 31 & -1,341930696378109 & 6,451869235304859 & 0,05482804889270922 & 0,5694170176403242 & 4,416924333519995 & 0,04212208612720639 & 0,5897125000677342 & 8,86521100790565 & 0,0866581794842557 & 0,5311744230227635 \\
 32 & -10,16467121571373 & 6,705508349541454 & 0,0934749698077619 & 0,4420693171728495 & 2,858371131410651 & 0,03722557506068529 & 0,4216580500297413 & 6,670274811535954 & 0,0977332470946321 & 0,538756085129185 \\
 33 & -1,107633246744663 & 7,390220519784453 & 0,07970661674181083 & 0,5081940602224835 & 3,815123656821275 & 0,04965529863397879 & 0,497023655525965 & 9,69945245311718 & 0,07039341131023045 & 0,4990744291774679 \\
 34 & 10,39451871866851 & 1,478170300165134 & 0,03602109509210016 & 0,5799113381048154 & 2,105782960173057 & 0,0861928382367878 & 0,4812528298377607 & 7,010226383015251 & 0,03535327789989237 & 0,4969989664577149 \\
 35 & -0,139925361351959 & 4,045190240200832 & 0,03579955552582782 & 0,5678362348598895 & 9,32386837735074 & 0,0858648976559044 & 0,5698752996855582 & 9,81895903799508 & 0,04704020956533079 & 0,5708692337579542 \\
 36 & 3,656124569605117 & 4,944179742822627 & 0,0931916525333049 & 0,4834640277977335 & 9,3847979419491 & 0,0840810063497973 & 0,4090254720704045 & 3,673880603434117 & 0,07158934023489803 & 0,5598196803643543 \\
 37 & 8,06734679112684 & 2,988178201634472 & 0,07482186947515447 & 0,4950144569660614 & 1,910770315838324 & 0,03877152903045122 & 0,5786659568499292 & 4,285706479098184 & 0,05940487550770145 & 0,4844513368813702 \\
 38 & 0,3011566621835722 & 8,78996207364355 & 0,04833452393773747 & 0,459203937360375 & 9,93467757777873 & 0,0973964303071365 & 0,4280349684453508 & 4,931948094146255 & 0,0969410287755626 & 0,4560660353529726 \\
 39 & -0,2469178958324283 & 6,242957525364009 & 0,0989568314234112 & 0,4205662570527381 & 4,73983230004983 & 0,0450110225161678 & 0,559400017887981 & 9,11978947943194 & 0,06189826031162779 & 0,493100819449519 \\
 40 & -5,704905642148452 & 4,447521595114223 & 0,06770523205457797 & 0,5047965897298675 & 3,987888630073714 & 0,03595136615254331 & 0,5929011044925536 & 5,22121823715344 & 0,04029431905467698 & 0,514129411681418 \\
 41 & 0,1425795779687519 & 6,919606258393273 & 0,0821756368975312 & 0,5976799418614426 & 8,04099037018168 & 0,03129262110881438 & 0,4686530881925831 & 3,421695873368174 & 0,05166215779587051 & 0,4208560803818513 \\
 42 & 0,6843702894502789 & 8,27200587323167 & 0,0547013388431661 & 0,4453254786423237 & 7,433883810628654 & 0,04612202343826998 & 0,5269730370279639 & 8,3641619521824 & 0,07996879390804805 & 0,5036755718396144 \\
 43 & -14,06571757369153 & 6,220712308892191 & 0,04765956545437692 & 0,4898959293071353 & 8,72249431241257 & 0,05059967705235588 & 0,4421274052875277 & 2,416276423317733 & 0,07373660930828208 & 0,5403850652535799 \\
 44 & 8,32346109443768 & 9,29716067662943 & 0,0656810603414827 & 0,4960066859011466 & 6,796263459296078 & 0,04037116107097755 & 0,4389885737869158 & 2,206183147865627 & 0,0804910989081056 & 0,5806221323132373 \\
 45 & -5,252317079780442 & 3,424683176707511 & 0,0867529985575643 & 0,4322548560407479 & 6,859255296971943 & 0,0955962752949951 & 0,517620399931194 & 5,826144942819697 & 0,0910408442631135 & 0,4167014408128501 \\
 46 & 6,18641388898792 & 3,789578783769164 & 0,07617326304382499 & 0,5418490045761073 & 1,363177133649232 & 0,03197249245240319 & 0,5284999992906743 & 4,9879001955338 & 0,04126125322920861 & 0,4895167206754735 \\
 47 & 0,4821906326173532 & 1,195769022526768 & 0,04928749082970199 & 0,5286859472378347 & 8,76714063033632 & 0,0870852412295278 & 0,5989638152903285 & 2,489565114307663 & 0,05568685410478115 & 0,5787056957942209 \\
 48 & 5,707663438868804 & 4,058614230578957 & 0,04743875933294599 & 0,4981149890354436 & 1,015171237842736 & 0,085826684646538 & 0,5470349320774566 & 6,343764732282688 & 0,0878769436185823 & 0,5449452367497044 \\
 49 & -0,1857958679022612 & 5,159763617905121 & 0,0850570422587659 & 0,5311491284465588 & 7,238990148015965 & 0,05867048711393652 & 0,5662410781222187 & 8,70600609334284 & 0,04088387548046445 & 0,5305143392409284 \\
 50 & 2,321594443893878 & 3,319181325761367 & 0,05391080867981007 & 0,5914435434600773 & 9,11776045377491 & 0,03583354001918901 & 0,4669413688494417 & 6,117669336479262 & 0,04553014156422459 & 0,4131324263527175 \\
 51 & 2,451069994722774 & 4,005247740100538 & 0,07620861067494449 & 0,5759225433332955 & 6,743529978394792 & 0,0967347280360008 & 0,4980631985926895 & 5,505849306951131 & 0,0971440363586883 & 0,45173467721102 \\
 52 & -0,961350897030818 & 7,710615861039985 & 0,07131617125307138 & 0,4007362522722102 & 3,372061685311945 & 0,04579335436967417 & 0,4845503268496217 & 8,6021995929235 & 0,0634681851083368 & 0,4207242810824543 \\
 53 & -10,67971490671366 & 4,728468083127391 & 0,06509200806253813 & 0,4959400801351127 & 9,81266481664175 & 0,0951346706530719 & 0,5047212015277842 & 3,180268423645401 & 0,06891913385037023 & 0,5758580789792047 \\
 54 & 1,375998039323297 & 2,740626958875014 & 0,06438323836566841 & 0,5077158168564866 & 1,813952930718781 & 0,0845523320869031 & 0,407247659218711 & 1,054506345314588 & 0,0998274789144316 & 0,5956344420441539 \\
 55 & -4,330863793299111 & 3,209443235555092 & 0,05150806767094947 & 0,5932432195310449 & 4,682408324655706 & 0,03641376226612134 & 0,4491940043173043 & 9,57800470422746 & 0,089704223465467 & 0,4702887735348434 \\
 56 & 6,961655278126058 & 3,013659237176324 & 0,0802684211883349 & 0,5733460791404235 & 7,441047648005995 & 0,0924462539370359 & 0,5216671466381739 & 8,15403131934562 & 0,05611934350568269 & 0,4910483719005932 \\
 57 & -5,30540083281513 & 2,088882156026381 & 0,07464091227469428 & 0,4740072756248339 & 4,228694369663863 & 0,03993178348974601 & 0,4402946238148492 & 4,912298433960689 & 0,06528359492264517 & 0,4923580906843156 \\
 58 & 3,395104400983763 & 7,9733680986292 & 0,0911843583955236 & 0,5314007871406556 & 6,375772536276909 & 0,0933026364108379 & 0,447244215655568 & 7,387561983870292 & 0,0977643384454912 & 0,479824590293356 \\
 59 & 1,330224789015745 & 5,105339790874075 & 0,06331718284342962 & 0,5503804706506275 & 9,5740821641842 & 0,07463500644132846 & 0,5814481952546582 & 5,810342341443878 & 0,03327635127803191 & 0,5550216356360829 \\
 60 & -0,2920604955263079 & 6,773414983805557 & 0,07595819905259191 & 0,5210834626114553 & 4,229974804881476 & 0,03532177545633547 & 0,5725615130215063 & 8,88452823254966 & 0,0932985710685446 & 0,5046388124857028 \\
 61 & -26,41840990525562 & 6,41701161854283 & 0,06246944646172664 & 0,4258923441426071 & 1,379491105318868 & 0,06306367927818108 & 0,4531443030507202 & 2,189456974183496 & 0,098078005933584 & 0,5174384564079299 \\
 62 & 6,630393088825135 & 6,476325542442256 & 0,07599435647188182 & 0,4345528705798566 & 6,969363897694016 & 0,06669302037176108 & 0,517427894398595 & 3,961825639220393 & 0,0929115322244988 & 0,4621043570336795 \\
 63 & 3,065748817227897 & 1,300093759156537 & 0,06741885075599173 & 0,4866345116967513 & 6,942052331590217 & 0,07107963920756023 & 0,4930672781311588 & 7,834870626014938 & 0,06446115603870624 & 0,5367219218216266 \\
 64 & -1,743105697039855 & 3,145607697855832 & 0,0415292992761004 & 0,5859414343226601 & 5,945295724435049 & 0,07414379406386953 & 0,4548629476738681 & 1,062923665624632 & 0,07636214080338054 & 0,4330291254331895 \\
 65 & -1,631255302904965 & 3,121291975208031 & 0,03133561741752311 & 0,524510056949801 & 8,03118879837252 & 0,0987376029412809 & 0,5128559599511593 & 7,572783963151993 & 0,0936192481138518 & 0,4523240987335903 \\
 66 & -9,50982692642515 & 7,367027947839741 & 0,07398108515426552 & 0,4753257600466925 & 1,589160147317211 & 0,07398588942349949 & 0,4644105814910987 & 1,719981611302808 & 0,04986976063479617 & 0,5394709698394433 \\
 67 & -4,496736421068983 & 8,56498650586069 & 0,0870950155799119 & 0,4807004710871501 & 8,85815411069566 & 0,0936690521145114 & 0,4120764594786555 & 2,422694718173737 & 0,05651923264880363 & 0,4832351547748815 \\
 68 & -1,002977917543076 & 9,43121814362608 & 0,03689894383172801 & 0,4801850915018365 & 6,762463110315146 & 0,0977990134875495 & 0,5880377299069773 & 8,48288777357423 & 0,05983735955021298 & 0,4837560035001434 \\
 69 & -9,23106545170694 & 8,47929696159459 & 0,05472141225455761 & 0,4949989702352598 & 4,776138672418544 & 0,03835162624659654 & 0,5785166211495689 & 4,459069855546714 & 0,06892772167363432 & 0,5908529311844791 \\
 70 & 14,99926304831282 & 4,708917711658332 & 0,0907364284065382 & 0,4550033334959929 & 5,373320800648651 & 0,03739924536269347 & 0,4468132406615605 & 1,831662246071428 & 0,05825558914206748 & 0,566434199642786 \\
 71 & 4,189181104505013 & 7,583635162152007 & 0,0936829708421792 & 0,4110679099121856 & 6,09002215219153 & 0,05895044789139901 & 0,5451253692040614 & 6,851478307492325 & 0,04763840334231468 & 0,5010633160850224 \\
 72 & 8,32837188138842 & 3,414315967033806 & 0,05054048464274665 & 0,5043307410524706 & 3,070996567281799 & 0,06968276328974509 & 0,5702669702356848 & 5,360783854112533 & 0,0388888383066967 & 0,5784155467566237 \\
 73 & -21,84200888491582 & 1,625552430388586 & 0,04546236831654488 & 0,5426078576324179 & 4,091176923293506 & 0,03836885039652842 & 0,504879512949007 & 4,527748284314512 & 0,05975894102927981 & 0,4670598506496871 \\
 74 & 6,604738977945651 & 3,42187897931262 & 0,0894449950697187 & 0,4941485628974172 & 4,96749741331263 & 0,06031151123783582 & 0,4119028637456998 & 4,581927607849956 & 0,0871948963692279 & 0,4176805754532423 \\
 75 & 15,27988423246628 & 6,229538426957454 & 0,0821521091215863 & 0,4118748449252315 & 2,377028497509498 & 0,0963110933106458 & 0,4746736556808109 & 5,398545541594487 & 0,0822574656780643 & 0,5258947233190487 \\
 76 & 1,377130763885458 & 9,15176346333775 & 0,04818682488618507 & 0,574408949666749 & 2,366656226877287 & 0,05059199769749539 & 0,4731074949729489 & 4,34290365370739 & 0,07492904049493838 & 0,4038656375892337 \\
 77 & -0,7756959351484244 & 8,1525112801581 & 0,0812975344211981 & 0,5112853751964107 & 4,530310672134927 & 0,06652912486134806 & 0,5946245231853162 & 1,568860626254933 & 0,04256882025605258 & 0,5491027801043556 \\
 78 & -9,50432792804336 & 4,670517118870347 & 0,04398267799553492 & 0,4519472673718893 & 2,155276187476066 & 0,0489411436180854 & 0,5694756278652033 & 1,947130206816878 & 0,06321309041201674 & 0,4122312147733568 \\
 79 & 3,754137123031758 & 6,801389212739803 & 0,07003309342623426 & 0,5129214842481318 & 9,92433910143913 & 0,0931392820982872 & 0,4104016495146404 & 5,23036608102518 & 0,0919529182014262 & 0,4465294974738288 \\
 80 & -4,499804654067136 & 4,406584351091798 & 0,0970211159011626 & 0,4772975938521527 & 2,270764431848441 & 0,0853562757122116 & 0,4669596920988839 & 4,42557413334857 & 0,0517800509473799 & 0,5645191809121128 \\
 81 & -13,9999061688152 & 8,30408622139241 & 0,0903948738093222 & 0,5053209583744364 & 2,454354187669313 & 0,0857681690400504 & 0,576877467653029 & 7,167289037979035 & 0,0831206320171505 & 0,4221339680792799 \\
 82 & -4,878405649764898 & 7,975675712689654 & 0,0809649515974139 & 0,5831990158889571 & 1,696376490191135 & 0,06783046119270637 & 0,5788560723486603 & 5,118512758012828 & 0,06554617006512321 & 0,5571586558825978 \\
 83 & -8,70611662690158 & 3,078380132262833 & 0,06327626752293522 & 0,4017988981679418 & 2,138629261588504 & 0,05374733302715019 & 0,5432443229891937 & 7,596013793576791 & 0,0999924467044729 & 0,5115245622576321 \\
 84 & 3,578654662113496 & 4,294915349281119 & 0,0664706641817281 & 0,5566289165852036 & 3,387516420897519 & 0,06654228293431248 & 0,4304935498494862 & 8,43232828156106 & 0,06121769693920587 & 0,5527076848906498 \\
 85 & 0,0423259541153967 & 9,56936665353494 & 0,06825963823641594 & 0,5532353068476618 & 7,233942984150266 & 0,07736032999103362 & 0,4274316541541941 & 8,96022693099 & 0,03996347733060968 & 0,5091073130786462 \\
 86 & -6,387446031119123 & 5,052123646744015 & 0,05306973453715137 & 0,4657297418468141 & 1,261670338009784 & 0,05392235260994765 & 0,4355571344074666 & 9,81884128475822 & 0,0434164193614162 & 0,546118275633825 \\
 87 & 3,029693711282459 & 1,767169254935782 & 0,0386273054889003 & 0,4174202422881018 & 9,99189882836594 & 0,0347107535224807 & 0,5316587974869709 & 4,794411329362948 & 0,04866392717198935 & 0,5202815575942617 \\
 88 & -0,791031105864576 & 8,38082427065573 & 0,0552127907704862 & 0,5821800194923659 & 3,864226559651998 & 0,05464216708052787 & 0,5051543239785165 & 4,445064096761643 & 0,03386907179361738 & 0,4659081177469465 \\
 89 & 2,97083681738158 & 9,22706363472046 & 0,06912140386549836 & 0,5630023378185795 & 2,929880196570098 & 0,07966659968684009 & 0,4992263633935188 & 6,920427033384797 & 0,03123407491202272 & 0,4077156601519876 \\
 90 & -7,110772604917624 & 3,106633667560681 & 0,04873990025387701 & 0,5697049535893381 & 8,30700777463754 & 0,05249501630989304 & 0,5669783762910612 & 1,13457393933062 & 0,07200760153994357 & 0,444672072474744 \\
 91 & -7,162305080703513 & 5,387619660361665 & 0,04581500385875453 & 0,5197835066282974 & 8,22551991773337 & 0,0831226693907787 & 0,4274987667560087 & 6,828948529150043 & 0,05943349593138421 & 0,5115861497137036 \\
 92 & 1,720368406482304 & 6,780548897314363 & 0,0870012111933092 & 0,5984908484842784 & 6,587571277412653 & 0,05694105237128053 & 0,5226697825342788 & 5,99005641944593 & 0,07837921322316456 & 0,4668236040752577 \\
 93 & 0,1494884045649762 & 8,09818524171748 & 0,0976595255928496 & 0,4848535596042637 & 2,153483785757176 & 0,04734211110674212 & 0,4077222220507315 & 7,143110762448606 & 0,04682210332386052 & 0,509084643832387 \\
 94 & -0,962821097034532 & 7,260437643310942 & 0,05083396575279076 & 0,54745303715986 & 8,38162990306018 & 0,05791050943688269 & 0,5794395625497382 & 2,722302743095399 & 0,083744615781603 & 0,5395428417251624 \\
 95 & 0,935122045036145 & 2,31788301454997 & 0,05976249487823031 & 0,5809204376238206 & 8,19172407354263 & 0,06806778195768599 & 0,5479374417815378 & 9,93990336225375 & 0,04266977230181523 & 0,5959901705863791 \\
 96 & 0,4328277236540677 & 4,113414259206836 & 0,07988404513340022 & 0,5262079910925557 & 7,436994191324823 & 0,07641699196144898 & 0,4316282085515486 & 9,59671807016201 & 0,07198966517052108 & 0,4563024970288369 \\
 97 & 0,853043516035702 & 3,229317247568396 & 0,04664288308967379 & 0,4916641780174803 & 9,19533615142581 & 0,0944216012526871 & 0,5792638413577232 & 6,980966479991427 & 0,07647312161060935 & 0,5407763967437385 \\
 98 & -3,516191871248886 & 9,20769534239692 & 0,07214599607965883 & 0,4111312070946356 & 7,568725646832963 & 0,0806902033322297 & 0,4952042950743076 & 8,94919747461644 & 0,04263443777290076 & 0,5735261960865083 \\
 99 & 19,67243760549639 & 2,270663590855083 & 0,0971170179967765 & 0,4613017967987272 & 8,22480560472966 & 0,0536024772561282 & 0,4745303011865644 & 3,970009703387312 & 0,0982032349257953 & 0,5029128207513967 \\
 100 & 1,544060514744405 & 4,017102041449901 & 0,04437090645089548 & 0,5073655615301741 & 3,724036679477685 & 0,07622772645899346 & 0,4368882837991222 & 5,350822342112657 & 0,07659119397629786 & 0,4056243763991382
\\ \bottomrule[0.1em]
	\end{tabular}
}
	\caption{All the necessary information to construct our MLA, trained from 100 densities with $N=1$ on a grid of 500 points,
	with $\lambda = 12 \times 10^{14}$ and $\sigma = 43$.
	For purposes of saving space, we do not list these densities. They may be reconstructed from these potentials via Numerov's method.}
	\label{tbl:MLAfunc}
\end{table}

%%%%%%%%%
%\renewcommand*\arraystretch{1.05}
%\input suppinfotbl2
%\clearpage

%%%%%%%%%

%%%%%%%%%
\label{page:end}